\documentclass[prb,twocolumn,floatfix,showpacs,superscriptaddress]{revtex4}
\usepackage{amsfonts}
\usepackage{amsmath}
\usepackage{amssymb}
\usepackage{graphicx}
\usepackage{natbib}
\usepackage[colorlinks=true]{hyperref}

\begin{document}

\title{c-axis resistivity, pseudogap, superconductivity and Widom line in doped Mott insulators}
\author{G. Sordi}
\affiliation{Theory Group, Institut Laue Langevin, 6 rue Jules Horowitz, 38042 Grenoble Cedex, France}
\author{P. S\'emon}
\affiliation{D\'epartement de physique and Regroupement qu\'eb\'equois sur les mat\'eriaux de pointe, Universit\'e de Sherbrooke, Sherbrooke, Qu\'ebec, Canada J1K 2R1}
\author{K. Haule}
\affiliation{Department of Physics \& Astronomy, Rutgers University, Piscataway, NJ 08854-8019, USA}
\author{A.-M. S. Tremblay}
\affiliation{D\'epartement de physique and Regroupement qu\'eb\'equois sur les mat\'eriaux de pointe, Universit\'e de Sherbrooke, Sherbrooke, Qu\'ebec, Canada J1K 2R1}
\affiliation{Canadian Institute for Advanced Research, Toronto, Ontario, Canada, M5G 1Z8}

\date{\today}

\begin{abstract}
Layered doped Mott insulators, such as the cuprates, show unusual
temperature dependence of the resistivity. Intriguingly, the resistivity perpendicular to the CuO$_2$ planes,
$\rho_c(T)$, shows both metallic ($d\rho_c/dT > 0$) and semi-conducting ($d\rho_c/dT<0$) behavior. 
We shed light on this puzzle by calculating $\rho_c$ for the two-dimensional Hubbard model within plaquette cellular dynamical mean-field theory and strong-coupling continuous-time quantum Monte Carlo as the impurity solver.
The temperature, $T$, and doping, $\delta$, dependencies of $\rho_c$ are controlled by the first-order transition between pseudogap and correlated metal phases from which superconductivity can emerge.
On the large doping side of the transition $\rho_c(T)$ is metallic, while on the low-doping side $\rho_c(T)$ changes from metallic to semi-conducting behavior with decreasing $T$.
As a function of doping, the jump in $\rho_c$ across the first-order transition evolves into a sharp crossover at higher temperatures.
This crossover coincides with the pseudogap temperature $T^*$ in the single-particle density of states, the spin susceptibility and other observables. 
Such coincidence in crossovers is expected along the continuation of the first-order transition into the super-critical regime, called the Widom line. 
This implies that not only the dynamic and the thermodynamic properties but also the DC transport in the normal state are governed by the hidden first-order transition. 
$\rho_c(T)$ has a high-temperature quasi-linear regime where it can exceed the Mott-Ioffe-Regel limit and when it has a minimum it is nearly parallel to the Widom line.
\end{abstract}
\maketitle

The puzzling behavior of electrical resistivity has been at the center of the high-temperature superconductivity conundrum from the very beginning~\cite{husseyBOOK}.
Normal-state resistivity is highly anisotropic~\cite{Ito:1991, nakamuraPRB,takenakaPRB}: the out-of-plane $c$-axis resistivity $\rho_c(T)$ can be orders of magnitude larger than the in-plane resistivity $\rho_{ab}(T)$.
In addition, for values of doping where the cuprates become superconducting, the normal-state $\rho_{ab}(T)$ and $\rho_c(T)$ can show contrasting behaviors: the in-plane resistivity $\rho_{ab}(T)$ is metallic, while the out-of-plane $c$-axis resistivity $\rho_c(T)$ can be both metallic or non-metallic (semi-conducting).
The opening of the pseudogap can be detected by the deviation in both the in-plane and the $c$-axis resistivities from their linear-$T$ behavior at high temperatures~\cite{Daou:2010,OlivierQCP,raffy1}.

The behavior of the $c$-axis resistivity, its relation to superconductivity and to the pseudogap are important issues for the understanding of cuprates.
Diverse explanations for the interplane resistivity have been offered over the years.
They can be classified into two groups, depending on whether it is unconventional for the in-plane physics or for the interlayer coupling.
The first group includes mechanisms based on spin-charge separation~\cite{Anderson:1992,andersonBOOK}, fluctuations of the phase of the superconducting order parameter~\cite{IoffeMillis1} or in-plane strong-coupling physics~\cite{michel-transport}.
The second group contains models where the interlayer tunnelling is influenced by disorder~\cite{RojoLevin}, by bosons~\cite{Alexandrov:1996,GutmanPRL,HoScho} or by interplane and in-plane charge fluctuations~\cite{turlakov}.

Here we examine the interplay between c-axis transport, superconductivity and pseudogap by studying the DC $c$-axis resistivity of a hole-doped Mott insulator represented by the one-band Hubbard model.
We explore the possibility that Mott physics --essentially the blocking of charge motion driven by strong Coulomb repulsion-- can determine the intricate doping and temperature behavior of $\rho_c$.
In our approach, the interplane transport is governed by the in-plane scattering, as in the first group of mechanisms for unconventional interplane conduction.

We solve the model using cellular dynamical mean-field theory (CDMFT)~\cite{kotliarRMP,maier} for a self-consistent $2\times 2$ plaquette. Recent developments in the algorithms~\cite{millisRMP} make the present study possible. 
Recent work~\cite{michel-transport} using a 2-site cluster strengthen the experimental correlation between the behavior of $\rho_c(T)$ and the opening of the pseudogap as revealed by angle-resolved photoemission (where the pseudogap appears as a lack of a quasiparticle peak at the antinode~\cite{DamascelliRMP}) and $c$-axis optical conductivity $\sigma_c(\omega)$ (where the pseudogap appears as a low-frequency suppression of spectral weight transferred to high frequencies~\cite{BasovRMP}).
Similar $c$-axis optical conductivity results have been shown in Ref.~\onlinecite{linPG} using an 8-site cluster.

A recent development requires that those findings be re-examined: in plaquette CDMFT, a first-order transition~\cite{sht,sht2} occurs at finite doping between a pseudogap and a correlated metal. 
This transition is connected to the Mott transition in the undoped model. The crossover to the pseudogap state, $T^*(\delta)$, lies along the thermodynamic crossover (known as Widom line~\cite{water1}) that begins at the critical endpoint of the transition and extends in the supercritical region. This indicates the common origin of the pseudogap and of the thermodynamic crossovers~\cite{ssht}.
The further step provided by the present work is to address the role of the Widom line for interplane transport. 
In agreement with previous studies~\cite{michel-transport,linPG}, we attribute the semi-conducting $\rho_c(T)$ to the development of the pseudogap. 
We further show that the temperature and doping dependence of $\rho_c$ is governed by the Widom line crossover generated by the pseudogap to correlated metal first-order transition, thereby providing a unified picture for explaining DC transport, thermodynamic and dynamic properties. The resistivity minimum, Ioffe-Regel limit, inflection points and their relation to the Widom line are also discussed.

{\it Method.}--
We consider the two dimensional Hubbard model on a square lattice,
\begin{equation}
  H=-\sum_{ij\sigma}t_{ij}c_{i\sigma}^\dagger c_{j\sigma}
  +U\sum_{i}\left(n_{i\uparrow }-\frac{1}{2}\right)\left(n_{i\downarrow }-\frac{1}{2}\right)
  -\mu\sum_{i\sigma} n_{i\sigma}
\label{eq:HM}
\end{equation}
with $t_{ij}$ the nearest neighbor hopping amplitude, $\mu$ the chemical potential and $U$ the on-site Coulomb repulsion.
The operators $c^{+}_{i\sigma}$ and $c_{i\sigma}$ respectively create and annihilate an electron with spin $\sigma$ at site $i$ and $n_{i\sigma}=c^{+}_{i\sigma}c_{i\sigma}$ is the number operator.
We solve this model using CDMFT~\cite{kotliarRMP,maier,tremblayR}.
In this approach, a cluster of lattice sites, here a $2\times2$ plaquette, is embedded in a self-consistent bath of non-interacting electrons.
The action of the cluster coupled to the bath is given by
\begin{equation}
  S = S_{c} +\int_{0}^{\beta} d\tau \int_{0}^{\beta} d\tau' {\bf \psi^{\dag}}(\tau) \hat{\Delta}(\tau,\tau') {\bf \psi}(\tau') ,
\label{eq:action}
\end{equation}
with $S_{c}$ the action of the cluster and $\hat{\Delta}$ the hybridization matrix.
We can then self-consistently determine $\hat{\Delta}$ by requiring that the infinite lattice and the plaquette have the same self-energy and the same Green's function on the plaquette:
$\hat{\Delta}(i\omega_{n}) =  i\omega_{n} +\mu -\hat{t}_{c}-\hat{\Sigma}_{c}(i\omega_{n}) -\hat{G}(i\omega_n)^{-1}$.
Here $\hat{t}_{c}$ is the cluster hopping, $\hat{\Sigma}_{c}$ is the cluster self-energy, and $\hat{G}(i\omega_n)=\sum_{\tilde{k}} \frac{1}{i\omega_{n} +\mu -\hat{t}(\tilde{k}) -\hat{\Sigma}_{c}(i\omega_{n})}$, with $\tilde{k}$ the superlattice wave vector.
To solve the impurity problem Eq.~(\ref{eq:action}), we use a continuous-time quantum Monte Carlo method~\cite{millisRMP,WernerCTQMC,hauleCTQMC} which sums all diagrams obtained by the expansion of the action Eq.~(\ref{eq:action}) in the hybridization $\hat{\Delta}$. With this approach, we constructed the normal state phase diagram of the model in Refs.~\onlinecite{sht,sht2,ssht}, and the superconducting phase diagram in Ref.~\onlinecite{sshtSC}.

Here we consider the $c$-axis DC electrical conductivity. We work in the limit where the hopping $t_{\perp}$ along the $c$-axis, i.e. between planes, is much smaller than in-plane energy scales, $t$ and $U$. To second order in $t_{\perp}$ then, the Green's functions entering the Kubo formula~\cite{MahanBOOK} for the conductivity can be evaluated to zero'th order in $t_{\perp}$ since each one of the vertices is already of order $t_{\perp}$. In addition, since $t_{\perp}$ at the vertices create an electron-hole pair where the electron and the hole lie on different planes, they do not interact since the Coulomb repulsion $U$ is local. Therefore, to leading order in $t_\perp$ there is no vertex correction and the $c$-axis conductivity is given by the bubble diagram
\begin{equation}
\begin{split}
\sigma_c(\Omega) =& 2\sigma_c^0 \int d\omega \frac{f(\omega)-f(\omega+\Omega)}{\Omega} \sum_{\bf k} t_{\perp}^2({\bf k}) A({\bf k}, \omega) \\ & A({\bf k}, \omega+\Omega).
\end{split}
\end{equation}
The factor 2 in front comes from the spin summation, $\sigma_c^0=e^2c/\hbar ab$, with $a,b$ in-plane and $c$ interplane lattice constants, $f(\omega)$ is the Fermi function, A$({\bf k},\omega)=-\frac{1}{\pi}$Im$G({\bf k},\omega)$ is the one-particle spectral function, and $t_\perp({\bf k}) = t_0 ( \cos k_x -\cos k_y )^2$ is the interplane hopping for collinear CuO$_2$ planes~\cite{AndersonTperp,Andersen:1994,Novikov:1993}. Clearly, the ${\bf k}$ dependence of $t_\perp({\bf k})$ makes the $c$-axis conductivity mainly sensitive to the antinodes, where the pseudogap develops~\cite{kyung}.  

To compute the DC conductivity, one has to take the limit $\Omega\rightarrow 0$.
We further simplify $\frac{f(\omega)-f(\omega+\Omega)}{\Omega}\rightarrow -\frac{df(\omega)}{d\omega} \rightarrow \delta(\omega)$ to obtain $\sigma_c(\Omega=0) =  2\sigma_c^0 \sum_{\bf k} t_{\perp}^2({\bf k}) A^2({\bf k},\omega=0)$. 
This expression is convenient because a simple quadratic extrapolation of the Matsubara data for the Green's function towards $\omega=0$ suffices to obtain the spectral weight at the Fermi energy. This expression for the conductivity is a good approximation if the spectral function does not vary much for $|\omega|\le T$. We have verified that this is a good approximation. 
The last step is to construct the lattice $A({\bf k},\omega=0)$ from cluster quantities.
There are different methods to do this interpolation. Here we use the Green's function periodization~\cite{davidGperiod}, which follows directly from the Self-Energy Functional Approach~\cite{potthoffSFA,tremblayR}.
We checked that cumulant periodization~\cite{stanescuK} gives qualitatively similar results.

We consider the following convenient model parameters: $U=6.2t$, which is larger than the critical value of the Mott transition at half-filling, $U_{\rm MIT}\approx 5.95$~\cite{phk,sht}, $t=1$, $t_0=0.05t$ and we present $\rho_c$ data in units of $\rho_c^0=1/\sigma_c^0$. 
At larger values of $U$, the critical point of the first-order transition moves to dopings that are larger and more consistent with cuprate phenomenology, but it is then located at temperatures too low to be accessible numerically~\cite{sht2}. 
To convert into physical units we use $t=0.35$eV and $c/ab=0.5\AA^{-1}$.

%
\begin{figure}
\centering{
\includegraphics[width=0.95\linewidth,clip=]{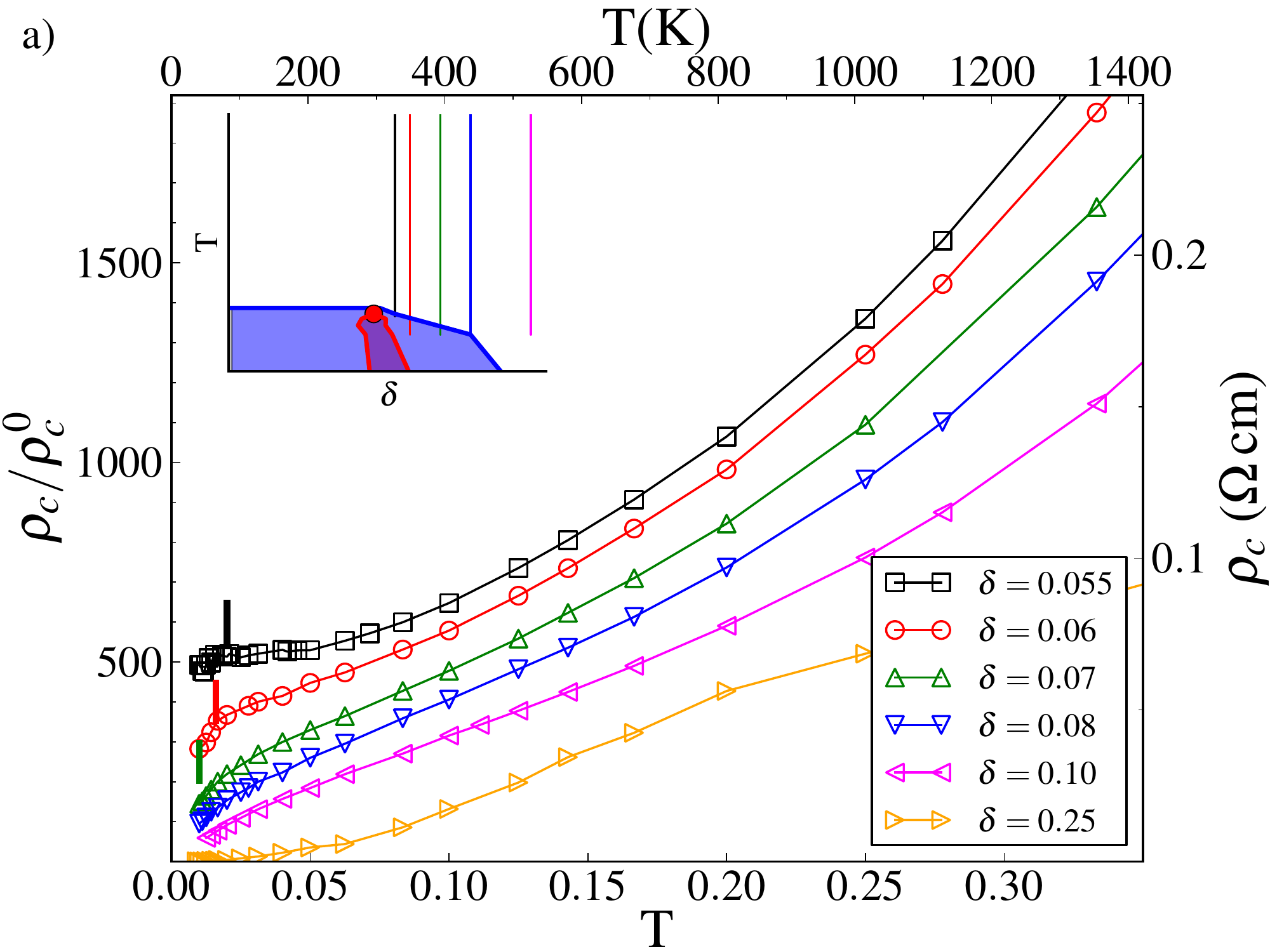}

\includegraphics[width=0.95\linewidth,clip=]{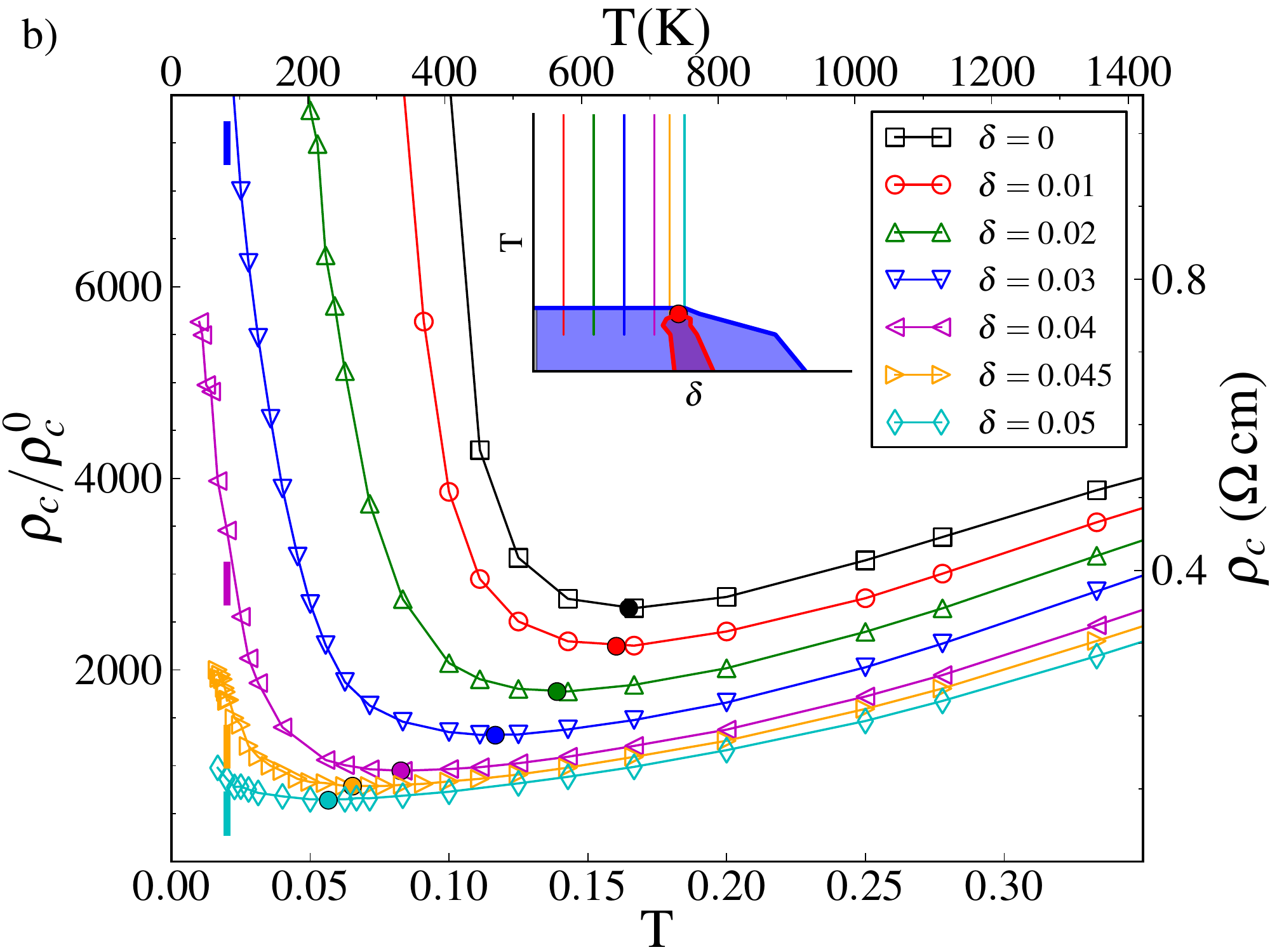}
}
\caption{(Color online) $c$-axis resistivity $\rho_c/\rho_c^0$ versus temperature $T$ for several values of hole doping $\delta=1-n$.
Data are obtained within CDMFT on a $2\times2$ plaquette in the normal state for $U=6.2$. Thick segments mark the superconducting transition temperature $T_c^d$.
Inset: $(\delta,T)$ phase diagram. A first-order transition (red/dark grey region) between a pseudogap and a correlated metal terminates at a critical point (red dot). It lies below the superconducting phase (blue/light grey area). Vertical lines indicate the values of doping of the resistivity data in the main panels.
a) For dopings above the first-order transition, $\rho_c(T)$ decreases monotonically with $T$.
b) For dopings below the first-order transition, $\rho_c(T)$ has a non-monotonic behavior. Filled dots mark the resistivity minimum.
}
\label{fig1}
\end{figure}
{\it Temperature dependence of $c$-axis resistivity.}--
Figure~\ref{fig1} shows the normal state $c$-axis resistivity $\rho_c$ as a function of temperature $T$ for different values of hole doping $\delta$.
The insets show the computed superconducting mean-field temperature $T_c^d$, below which  Cooper pairs form within the $2\times2$ plaquette~\cite{sshtSC,gullSC} if we do not allow for antiferromagnetism. Buried below the superconducting phase (blue/light grey region), there is  the first-order transition from a pseudogap state to a correlated metal terminating in a critical point at finite $T$ and finite $\delta$ (red/dark grey region).

We study $\rho_c(T)$ in a broad range of temperature: at high $T$, $\rho_c(T)$ has an approximate linear behavior with no sign of saturation.
It can exceed the generalized Mott-Ioffe-Regel maximum metallic resistivity saturation limit for anisotropic systems, $\rho_{ab}\rho_c \le \rho_0^2$. This has also been observed in DMFT for isotropic systems, as discussed recently~\cite{Deng:2012,vlad,vlad_qwl}. 
At low $T$, CDMFT allows us to study the underlying normal state of the model by suppressing the superconducting order. Hence, we can reveal the normal state $\rho_c$ below $T_c^d$.

The overall magnitude of $\rho_c$ decreases with increasing doping and presents two characteristic behaviors.
(1) For values of doping larger than the first-order transition (Fig.~\ref{fig1}a), $\rho_c$ increases monotonically with $T$, i.e. has a metallic behavior, $d\rho_c/dT>0$.
Moving away from the first-order transition, the zero-temperature intercept $\rho_c(T\rightarrow 0)$ decreases, and in the low-$T$ regime $\rho_c(T)$ gradually approaches Fermi liquid behavior (see $\delta=0.25$).
(2) For values of doping smaller than the first-order transition (Fig.~\ref{fig1}b), $\rho_c(T)$ has a non-monotonic behavior.
$\rho_c(T)$ is metallic at high temperatures, goes through a minimum at $T_{\rm min}$ (full circles) and crosses over to a semi-conducting $T$ dependence ($d\rho_c/dT<0$) as $T$ is decreased.
$T_{\rm min}$ decreases with increasing doping, leading to a wider range of $T$ with metallic dependence when $\delta$ is increased toward the first-order transition.

Why does $\rho_c(T)$ increase with decreasing $T$?
Many works have rooted the upturn in $\rho_c(T)$ to the onset of the pseudogap state~\cite{husseyBOOK,BasovRMP}.
Within cluster DMFT methods, previous work reported both the occurrence of a pseudogap state close to the Mott insulator~\cite{kyung,tremblayR,michelCFR,ssht}, and the correlation between the semi-conducting-like $\rho_c(T)$ behavior and the pseudogap phase~\cite{michel-transport}.
Our systematic data in Fig.~\ref{fig1} confirm these findings, but recast these results in a new framework: the first-order transition between the pseudogap and the correlated metal defines a watershed for transport properties. It separates a regime where $\rho_c(T)$ shows a metallic behavior from a regime where $\rho_c(T)$ has a non-monotonic behavior.

\begin{figure}
\centering{
\includegraphics[width=0.95\linewidth,clip=]{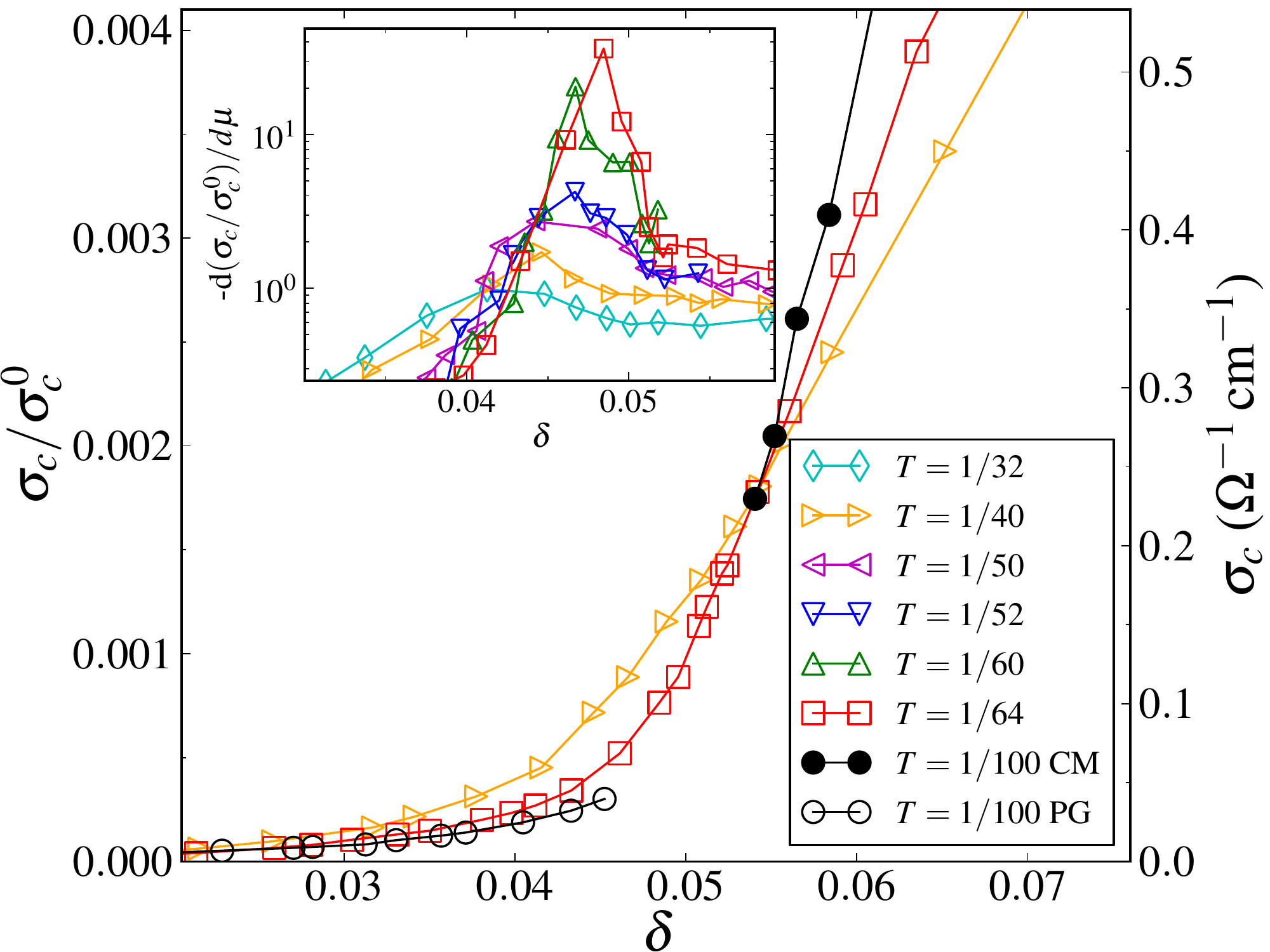}
}
\caption{(Color online) Normal-state $c$-axis conductivity $\sigma_c/\sigma_c^0$ versus doping.
For $T<T_p$, where $T_p$ is the temperature of the second-order critical endpoint, $\sigma_c(\delta)$ has a discontinuous jump (line with circles).
Above $T_p$, $\sigma_c(\delta)$ is continuous and rapidly crosses over from a low interplane conductive state (the pseudogap, PG) to a high conductive state (the correlated metal, CM).
The inflection point in $\sigma_c(\mu)$ defines the characteristic crossover temperature $T_{\sigma_c} (\delta)$.
Inset: $-d(\sigma_c/\sigma_c^0)/d\mu$ versus $\delta$ in a semilogarithmic scale. The maximum of the peak is used to locate the inflection point in $\sigma_c(\mu)$.
}
\label{fig2}
\end{figure}

{\it Doping dependence of $c$-axis resistivity.}--
To further characterize how the first-order transition modifies the $c$-axis transport, we present in Fig.~\ref{fig2} the $c$-axis conductivity $\sigma_c$ as a function of doping for several temperatures.
For $T<T_p$,  $\sigma_c(\delta)$ is discontinuous at the transition with a jump of almost one decade of magnitude (lines with circles).
Coexistence is found for $\sigma_c(\mu)$ in the $(T,\mu)$ plane.
The jump in the conductivity is a hallmark of the first-order nature of the transition.

\begin{figure}
\centering{
\includegraphics[width=0.95\linewidth]{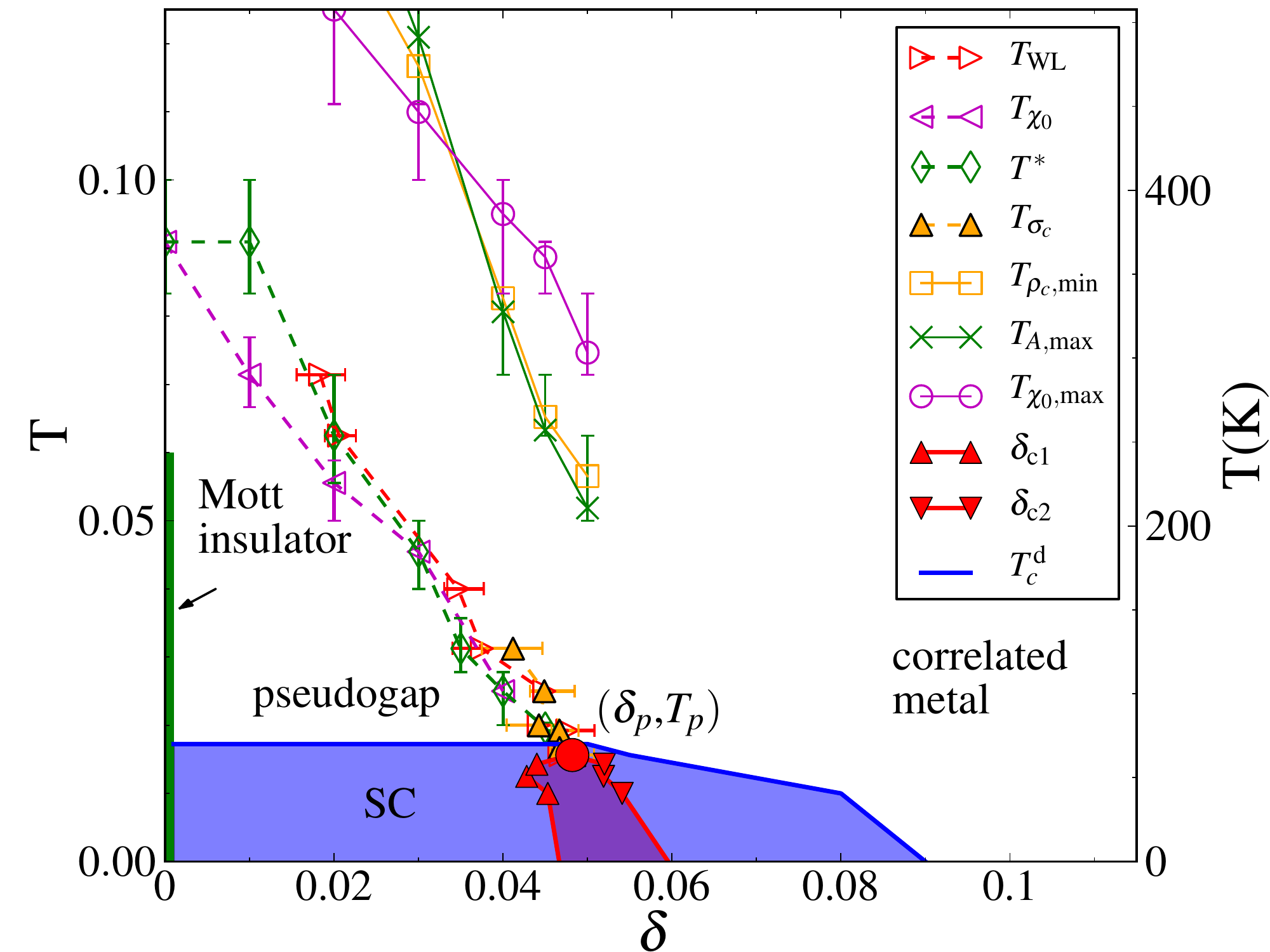}
}
\caption{(Color online) Temperature versus doping phase diagram of the two dimensional Hubbard model within plaquette CDMFT for $U=6.2$.
Below the superconducting region delineated by $T_c^d$ (blue/light grey area), the first-order transition (red/dark grey area) terminating at the critical endpoint $(\delta_p, T_p)$ (circle) separates a correlated metal from a pseudogap metal.
$T_{\sigma_c}(\delta)$ is the temperature where $\sigma_c(\mu)$ has an inflection point.
It follows $T^*$ and $T_{\rm WL}$, i.e. the dynamic and thermodynamic supercritical crossovers determined by the inflection in the local density of states $A(\omega=0,T)$ and in the charge compressibility $\kappa(\mu)$ respectively. 
The pseudogap scale can be identified also as inflection points in the local spin susceptibility $\chi_0(T)$, $T_{\chi_0}$. 
$T_{\rho_c, \rm min}$ is the temperature where $\rho_c(T)$ has a minimum. It scales with the temperature where $A(\omega=0,T)$ [$\chi_0(T)$] peaks, $T_{A, \rm max}$ [$T_{\chi_0, \rm max}$], and can be used as a predictor of the crossover $T_{\sigma_c}$.
}
\label{fig3}
\end{figure}

At the critical temperature $T_p$, the two distinct phases separated by the transition (pseudogap and correlated metal) merge into one.
$\sigma_c(\delta)$ is continuous with an infinite slope at $\delta_p$, $d\sigma_c/d\mu|_{\delta_p}\rightarrow \infty$.
The endpoint of the first-order transition generates a characteristic crossover in the supercritical region $T>T_{p}$: the divergence in $d\sigma_c/d\mu$ is replaced by a peak, whose maximum decreases away from $T<T_p$ (see inset of Fig.~\ref{fig2}).
From the conductivity data, we define a characteristic temperature $T_{\sigma_c}$ as the point at which the derivative $d\sigma_c/d\mu$ reaches its maximum. 
$T_{\sigma_c}(\delta)$ characterizes the pseudogap to correlated metal crossover. 
In Fig.~\ref{fig3} we show the crossover line $T_{\sigma_c}(\delta)$ (line with full orange triangles) that sharpens towards the first-order transition.
This crossover sticks out of the superconducting region and can be used to predict the first-order transition hidden by superconductivity.
The rapid rise of conductivity (or, equivalently, drop of resistivity) with doping above the superconducting region is a stringent prediction of our theory.

{\it Signature of the Widom line in the $c$-axis transport.}--
In a recent work~\cite{ssht} we have linked the {\it dynamic} crossover corresponding to the opening of the pseudogap in the density of states to the {\it thermodynamic} continuation of the first-order transition in the supercritical region, called Widom line.
Theoretically, the latter is defined as the line where the maxima of different thermodynamic response functions merge close to the critical endpoint~\cite{water1}.
The Widom-line crossover governs DC {\it transport} properties as well. This is shown Fig.~\ref{fig3} that compares three crossover lines: $T_{\sigma_c}$, i.e. the interplane transport crossover in Fig.~\ref{fig2}, $T^*$, i.e. the dynamic crossover signaling the pseudogap phase, obtained from the inflection point in the local density of states~\cite{ssht} $A(\omega=0)$ along paths at constant $\delta$, and $T_{\rm WL}$, i.e. the thermodynamic crossover identified by a peak in the charge compressibility $\kappa=1/n^2dn/d\mu$ (or 
by the inflection point in the $T$ dependent spin susceptibility~\cite{ssht} (Knight shift), $T_{\rm \chi_0}$).
The three phenomena are concomitant.
All crossover temperatures decrease with increasing doping and they end at the critical endpoint $(\delta_p, T_p)$.

Another important feature of $\rho_c(T)$ is the appearance of minima in Fig.~\ref{fig1}b. In Fig.~\ref{fig3} we also show the location of these minima $T_{\rho_c, \rm min}$. 
While $T_{\rho_c, \rm min}$ occurs at higher $T$ than $T_{\sigma_c}$, the lines are nearly parallel. Thus $T_{\rho_c, \rm min}$ can be used as a high temperature predictor of the crossover $T_{\sigma_c}$ (or, equivalently, $T^*$).
It also scales with $T_{A,\rm max}$, the maximum of the local density of states $A(\omega=0, T)$.
However, $T_{\rho_c, \rm min}$ does not end at the critical point $(\delta_p, T_p)$; the minimum of $\rho_c(T)$ becomes more shallow with increasing $\delta$ (see Fig.~\ref{fig1}b) and eventually disappears for values of doping larger than the first-order transition.
Therefore our systematic analysis rules out interpretations based on a linear extrapolation of $T_{\rho_c, \rm min}$ (or $T_{A, \rm min}$) to $T\rightarrow 0$.
This would lead to a value of critical doping $\delta\approx 0.07$, in contradiction with the metallic like $\rho_c(T)$ above $\delta\approx 0.05$.

$T_{\sigma_c}(\delta)$ intersects the superconducting phase delimited by $T_c^d$.
This result supports our discovery that within cluster DMFT, the pseudogap and superconductivity are distinct phenomena~\cite{sshtSC}, a result confirmed by larger cluster calculations~\cite{gullSC,sakaiSC}.
Here, we find that superconductivity can appear from a normal-state where out-of-plane conduction is semi-conducting like (i.e. non-metallic on the small $\delta$ side of the transition) or  metallic (on the large $\delta$ side of the transition).

In summary, the first-order transition ending at a critical point $(\delta_p, T_p)$ and its associated crossover in the supercritical region emerges as the unifying mechanism to interpret the out-of-plane transport.
It is a watershed separating a regime where $\rho_c(T)$ shows a metallic behavior from a regime where $\rho_c(T)$ has a non-monotonic behavior.
The rapid increase of $\sigma_c$ with doping coincides with the pseudogap temperature $T^*$ and with the Widom line, namely the thermodynamic crossover generated by the first-order transition in the supercritical region.
The resistivity minimum is distinct from $T^*$ but follows a line that is a large $T$ precursor.
Thus we ascribe DC transport, dynamic and thermodynamic crossovers to a common origin.

We acknowledge S. Allen for technical help and N. Doiron-Leyraud for useful discussion. This work was partially supported by FQRNT, by the Tier I Canada Research Chair Program (A.-M.S.T.), and by NSF DMR-0746395 (K.H.). Simulations were performed on computers provided by CFI, MELS, Calcul Qu\'ebec and Compute Canada.


\end{document}